\documentclass[10pt,conference]{IEEEtran}
\usepackage{graphicx}
\usepackage{amsmath}
\newtheorem{theorem}{Theorem}[section]

\newtheorem{corollary}[theorem]{Corollary}

\newcommand\abs[1]{\lvert#1\rvert}
\font\msbm=msbm10 at 10pt
\newcommand{\ZZ}{\mbox{\msbm Z}}
\newcommand{\FF}{\mbox{\msbm F}}
\newcommand{\RR}{\mbox{\msbm R}}

\newcommand{\TT}{\mbox{\msbm T}}
\font\msb=msbm10 at 8pt
\newcommand{\ZZZ}{\mbox{\msb Z}}
\font\msbb=msbm10 at 25pt
\newcommand{\Z}{\mbox{\msbb Z}}

\begin{document}

\title{Isometries and Binary Images of Linear Block Codes over $\Z_4+u\Z_4$ and $\Z_8+u\Z_8$ }
\author{
   \IEEEauthorblockN{Virgilio P. Sison and Monica N. Remillion}
   \IEEEauthorblockA{Institute of Mathematical Sciences and Physics\\
     University of the Philippines, Los Ba\~{n}os\\
     College, Laguna 4031, Philippines\\
     Email: \{vpsison, mnremillion\}@up.edu.ph}
 }
\maketitle
\begin{abstract}
\boldmath Let $\FF_2$ be the binary field and $\ZZ_{2^r}$ the residue class ring of integers modulo $2^r$, where $r$ is a positive integer. For the finite $16$-element commutative local Frobenius non-chain ring $\ZZ_4+u\ZZ_4$, where $u$ is nilpotent of index $2$, two weight functions are considered, namely the Lee weight and the homogeneous weight. With the appropriate application of these weights, isometric maps from $\ZZ_4+u\ZZ_4$ to the binary spaces $\FF_2^4$ and $\FF_2^8$, respectively, are established via the composition of other weight-based isometries. The classical Hamming weight is used on the binary space. The resulting isometries are then applied to linear block codes over $\ZZ_4+u\ZZ_4$ whose images are binary codes of predicted length, which may or may not be linear. Certain lower and upper bounds on the minimum distances of the binary images are also derived in terms of the parameters of the $\ZZ_4+u\ZZ_4$ codes. Several new codes and their images are constructed as illustrative examples. An analogous procedure is performed successfully on the ring $\ZZ_8+u\ZZ_8$, where $u^2=0$, which is a commutative local Frobenius non-chain ring of order 64. It turns out that the method is possible in general for the class of rings $\ZZ_{2^r}+u\ZZ_{2^r}$, where $u^2=0$, for any positive integer $r$, using the generalized Gray map from $\ZZ_{2^r}$ to $\FF_2^{2^{r-1}}$. 
\end{abstract}
\vskip .1in

\section{Introduction}
In this present work, we give much consideration to the fact that the finite commutative ring $\ZZ_{2^r}+u\ZZ_{2^r},u^2=0$ is a Frobenius ring. The reason is that Frobenius rings, which include the residue class rings $\ZZ_M$ and the Galois fields $\FF_q$, constitute the most general class of rings that admit a homogeneous weight, a far reaching generalization of the Lee metric over $\ZZ_4$. Moreover, since two of the classical theorems in coding theory, namely the Extension Theorem and the MacWilliams identities, generalize to the case of finite Frobenious rings, these rings are viewed as the most appropriate alphabet for code-theoretic purposes. Moreover, a homogeneous weight on $\ZZ_{2^r}+u\ZZ_{2^r}$ induces an isometry into its binary image. An isometry is basically a weight-preserving map. 

The material is organized as follows. Section~\ref{sect:pre} introduces the basic definitions and the relevant concepts such as Frobenius rings, homogeneous weights, linear block codes, and the structural properties of the alphabet ring. Section~\ref{sect:res} gives the main results of the paper, while Section~\ref{sect:con} summarizes the method used for a possible generalization. We refer the reader to \cite{woo} for a characterization of Frobenius rings. Greferath and Schmidt \cite{gre:sch:2} obtained an existence and characterization theorem for (left) homogeneous weights on arbitrary finite rings, and Honold \cite{hon} showed that a homogeneous weight on a finite Frobenius ring can be expressed in terms of its generating character [please see (\ref{cha}) below]. 

\section{Preliminaries and definitions}
\label{sect:pre}
\subsection{Frobenius rings and homogeneous weight}
Let $R$ be a finite ring with identity $1 \ne 0$, and $\TT$ be the multiplicative group of unit complex numbers. The group $\TT$ is a one-dimensional torus. A {\it character} of $R$ (considered as an additive abelian group) is a group homomorphism $\chi: R \rightarrow \TT$. The set of all characters $\widehat{R}$ (called the {\it character module of $R$}) is a right (resp. left) $R$-module whose group operation is pointwise multiplication of characters while scalar multiplication is given by $\chi^r(x) = \chi(rx)$ (resp. $^r\!\chi(x) = \chi(xr)$). A character $\chi$ of $R$ is called a {\it right (resp. left) generating character} if the mapping $\phi: R \rightarrow \widehat{R}$ given by $\phi(r) = \chi^r$ (resp. $\phi(r) = \: ^r\!\chi$) is an isomorphism of right (resp. left) $R$-modules. It is known that $\chi$ is a right (resp. left) generating character if and only if $\ker \chi$ contains no nonzero right (resp. left) ideals, and that for finite rings, $\chi$ is a right generating character if and only if it is a left generating character. We now call $R$ as {\it Frobenius} if $R$ admits a right (resp. left) generating character.  

Alternatively we say that $R$ is Frobenius if $Soc(R) \cong  R/J(R)$ as right (resp. left) $R$-modules. Here, $Soc(R)$ denotes the socle of $R$, the sum of all minimal (or simple) right (resp. left) ideals of $R$, while $J(R)$, the Jacobson radical of $R$, is the intersection of all maximal right (resp. left) ideals of $R$. In a commutative local ring the Jacobson radical is necessarily the unique maximal ideal.

A homogeneous weight on an arbitrary finite ring $R$ with identity is defined in the sense of \cite{gre:sch:2}. Let $\RR$ be the set of real numbers and $Rx$ be the principal (left) ideal generated by $x \in R$. A weight function $w:R \rightarrow \RR$ is called (left) homogeneous if the following conditions are satisfied.
	\begin{itemize}
 	\item [(i)] If $Rx = Ry$, then $w(x) = w(y)$ for all $x,y \in R$.
	\item [(ii)] There exists a real number $\Gamma \ge 0$ such that 
		\begin{equation}
		\sum_{y \in Rx} w(y) = \Gamma \cdot \abs{Rx}, \:\:  \text{for all} \:  x \in R \setminus \{0\} \: . \label{ave} 
        	\end{equation}
	\end{itemize}
	
Right homogeneous weights are defined accordingly. If a weight is both left homogeneous and right homogeneous, we call it simply as a homogeneous weight. The constant $\Gamma$ in (\ref{ave}) is called the {\it average value} of $w$. The homogeneous weight is said to be {\it normalized} if $\Gamma = 1$.  The weight $w$ is extended naturally to $R^n$, the free module of rank $n$ consisting of $n$-tuples of elements from $R$, via $w(z) = \sum_{i=0}^{n-1} w(z_i)$ for $z = (z_0,z_1,\ldots,z_{n-1}) \in R^n$. The homogeneous distance metric $\delta: R^n \times R^n \longrightarrow \RR$ is defined by $\delta(x,y)= w(x-y)$ for $x,y \in R^n$. 

It was proved in \cite{hon} that, if $R$ is Frobenius with generating character $\chi$, then every homogeneous weight $w$ on $R$ can be expressed in terms of $\chi$ as follows.
\begin{equation}\label{cha}
 w(x) = \Gamma \left[ 1 - \frac{1}{\abs{R^{\times}}} \sum_{v \in R^{\times}} \chi(xv)\right] 
\end{equation}
where $R^{\times}$ is the group of units of $R$. 

\subsection{Linear Block Codes}
A linear block code $C$ of length $n$ over a finite ring $R$ is a nonempty subset of $R^n$. The code $C$ is called {\it right (resp. left) $R$-linear} if $C$ is a right (resp. left) $R$-submodule of $R^n$. If  $C$ is both left $R$-linear and right $R$-linear, especially in the case when $R$ is commutative, we simply call $C$ a linear block code over $R$. A matrix $G \in R^{k \times n}$ is a {\it generator matrix} of a rate-$k/n$ linear block code $C$ if the $k$ rows of $G$ span $C$ and no proper subset of the rows of $G$ generates $C$, in which case, $C = \{v \in R^n \mid v=uG, u \in R^k \}$. If $C$ is a free module, then $C$ is called a {\it free code}. 

The minimum homogeneous distance $\delta_{\tt min}$ of $C$ is defined to be $\delta_{\tt min} = \min \lbrace \delta(x,y)\mid x,y\in C,x \neq y \rbrace$, while the minimum homogeneous weight $w_{\tt min}$ of $C$ is defined to be $w_{\tt min} = \min \lbrace w(x)\mid x\in C,x \neq 0 \rbrace$. For linear codes, the minimum distance is equal to the minimum weight.

\begin{figure}[!t]
\centering
\includegraphics[width=1in]{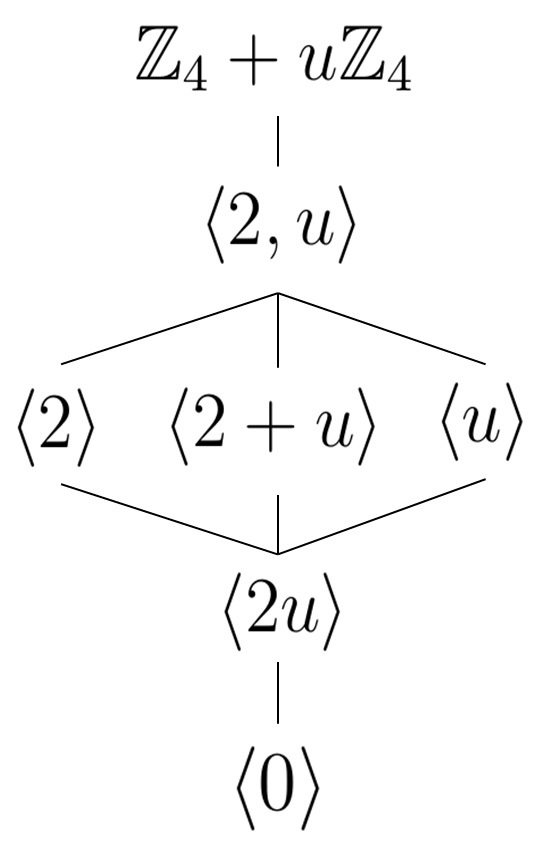}
%\resizebox{.13\textwidth}{!}{\includegraphics{z4.jpg} }
\caption{Lattice of Ideals of $\ZZ_4+u\ZZ_4$}
\label{fig:lat01}
\end{figure}

We begin with the algebraic structure of the code alphabet $R = \ZZ_4+u\ZZ_4 = \{a+ub \mid a,b \in \ZZ_4, u^2=0 \}$. Martinez-Moro and Szabo \cite{mor:sza} gave the 21 commutative local rings with identity of order 16. Of these, 15 are Frobenius, where 8 are chain rings and 7 are non-chain rings, as given in \cite{dou:sal:sza}. The ring $R$ belongs to the class of finite $16$-element Frobenius commutative rings with identity which are local, non-chain, and non-principal ideal rings. Its characteristic is $4$ and its additive group is $\ZZ_4 \times \ZZ_4$. The Jacobson radical is the unique maximal non-principal ideal $\langle 2,u \rangle = \{0,2,u,2u,3u,2+u,2+2u,2+3u \}$ which consists of the zero element and all the zero divisors of $R$. The remaining 8 elements in $R \setminus \langle 2,u \rangle$ form the multiplicative group of units of $R$ which can be described as $a+ub, a \in \{1,3\},$ the group of units of $\ZZ_4$. The lattice in Fig. \ref{fig:lat01} also shows the unique minimal principal ideal $\langle 2u \rangle = \{0,2u\}$ which is significant in deriving the homogeneous weight on $R$. This minimal ideal is the socle. Thus clearly, $R/J(R) \cong Soc(R) \cong \ZZ_2$ as $R$-modules. The generating character of $R$ is given by $\chi(a+ub) = e^{\frac{\pi i}{2}(a+b)} = i^{a+b}$. It can be shown that $R$ is isomorphic to the quotient ring $\ZZ_4[x]/\langle x^2 \rangle$.   

Yildiz and Karadeniz in \cite{yil:kar} emulated the Gray map $\ZZ_4 \rightarrow \FF_2^2$ to define the map $\phi_3:R \rightarrow \ZZ_4^2$ by $\phi_3(a+ub)=(b,a+b)$, and then extended the classical Lee weight on $\ZZ_4$ to $R$ by $w_L(a+ub) = w_L(b,a+b)=w_L(b)+w_L(a+b)$. However, this weight is not homogeneous.  

\section{Results and Discussion}
\label{sect:res}
\subsection{Linear block codes over $\ZZ_4+u\ZZ_4$}
With the structure of $R$ and its generating character known, from (\ref{cha}) we derive directly the explicit form of the homogeneous weight $w_{hom}$ on $R$.   
\begin{corollary}
The homogeneous weight on $\ZZ_4+u\ZZ_4$ is given by
\begin{equation} \label{hom4} w_{hom}(x) = \begin{cases}
                         0 & {\rm if} \ x=0 \\
                         2\Gamma & {\rm if }\ x=2u\\
												\Gamma & {\rm otherwise}\\
                    \end{cases}
\end{equation}
\end{corollary}
Observe that the highest possible homogeneous weight is given by the nonzero element of the minimal ideal, while all the other nonzero elements of $R$ take the average value $\Gamma$ as their common weight. Later we will choose $\Gamma =4$.

To obtain certain binary images, in this paper we apply the mapping $\phi_2$ from $\ZZ_4^2$ onto $\FF_2^4$ as a natural extension of the Gray map on $\ZZ_4$. The map $\phi_2$ is given by
\begin{equation}
\phi_2(x,y)=(x_2,x_1+x_2,y_2,y_1+y_2) 
\end{equation}
where $x_1+2x_2$ and $y_1+2y_2$ are the binary expansions of $x$ and $y$ respectively. We remark that $\phi_2$ is a non-linear bijection but is an isometry from $(\ZZ_4^2,\text{Lee weight})$ onto $\FF_2^4, \text{Hamming weight})$. Then we take the composition $\phi_2 \phi_3$ from $\ZZ_4+u\ZZ_4$ onto $\FF_2^4$ to get 
\begin{equation} 
\phi_2\phi_3(a+ub)=(b_2,b_1+b_2,c_2,c_1+c_2)
\end{equation}
where $b=b_1+2b_2$, $c=c_1+2c_2$ are the $2$-adic representations of $b$ and $c=a+b$ respectively. The following results are determined quite straightforwardly. 
\begin{theorem}
The map $\phi_2\phi_3$ is a weight preserving map from $(\ZZ_4+u\ZZ_4, \text{Lee weight})$ to $(\FF_2^4, \text{Hamming weight})$.
\end{theorem}
\begin{theorem}
Let $B$ be a linear block code over  $\ZZ_4+u\ZZ_4$ of length $n$, then the associated image $\phi_2\phi_3(B)$ is a (possibly nonlinear) binary block code of length $4n$. If $B$ is free of rank $k$, then necessarily the binary image has rank $4k$. 
\end{theorem}

We now proceed with determining an isometry, in terms of the homogeneous weight, from $R$ to the binary space $\FF_2^8$, which has twice the length of the binary image under $\phi_2 \phi_3$. Define the map $\phi_4$ from $R$ to $\FF_2^8$ by $\phi_4(a+ub)=(b_2,b_2+b_1,b_2+a_2,b_2+a_2+b_1,b_2+a_1,b_2+a_1+b_1,b_2+a_1+a_2,b_2+a_1+a_2+b_1)$ where $a=a_1+2a_2$ and $b=b_1+2b_2.$ 

Note that $\phi_4$ is a nonlinear non-surjective map. This seemingly convoluted map is in fact a composition of various isometries $R \rightarrow \FF_2^4 \rightarrow \ZZ_{16} \stackrel{\gamma_2}{\rightarrow} \FF_2^8$, where the last component $\gamma_2$ is the generalized Gray map given by Carlet \cite{car} as follows: let $x \in \ZZ_{16}$ where $x=x_1+2x_2+4x_3+8x_4$ is its binary expansion. The image of $x$ under $\gamma_2$ is the following Boolean function $f_x$ on $\FF_2^3$ given by  
\begin{table} [ht]
\begin{center}
\begin{tabular} {c c c c } 
$f_x:$ &$\FF_2^3$ & $\rightarrow$ & $\FF_2$ \\
			 &$(y_1,y_2,y_3)$&  $\mapsto$			&$x_4+x_1y_1+x_2y_2+x_3y_3$ \\
			 &$(0,0,0)$	 &    $\mapsto$					&$x_4$ \\
			 &$(0,0,1)	$ &    $\mapsto$					&$x_4+x_3$ \\
			 &$(0,1,0)	$ &    $\mapsto$					&$x_4+x_2$ \\
			 &$(0,1,1)	$ &    $\mapsto$					&$x_4+x_2+x_3$ \\
			 &$(1,0,0)	$ &    $\mapsto$					&$x_4+x_1$ \\
			 &$(1,0,1)	$ &    $\mapsto$					&$x_4+x_1+x_3$ \\
			 &$(1,1,0)	$ &    $\mapsto$					&$x_4+x_1+x_2$ \\
			 &$(1,1,1)	$ &    $\mapsto$					&$x_4+x_1+x_2+x_3$ \\
\end{tabular}
\end{center}
\end{table}

The main idea for this composition is to preserve in a systematic manner the isometry from $R$ to $\FF_2^8$ as seen in the next theorems.
\begin{theorem}
With $\Gamma=4$, the map $\phi_4$ is a weight preserving map from $(\ZZ_4+u\ZZ_4, \text{Homogeneous weight})$ to $(\FF_2^8, \text{Hamming weight})$.
\end{theorem}
\begin{theorem}
Let $B$ be a linear block code over  $\ZZ_4+u\ZZ_4$ of length $n$, then the associated image $\phi_4(B)$ is a (possibly nonlinear) binary block code of length $8n$. If $B$ is free of rank $k$, then necessarily the binary image has rank $8k$. 
\end{theorem}

\subsection{Linear block codes over $\ZZ_8+u\ZZ_8$}

\begin{figure}[!t]
\centering
\includegraphics[width=1.85in]{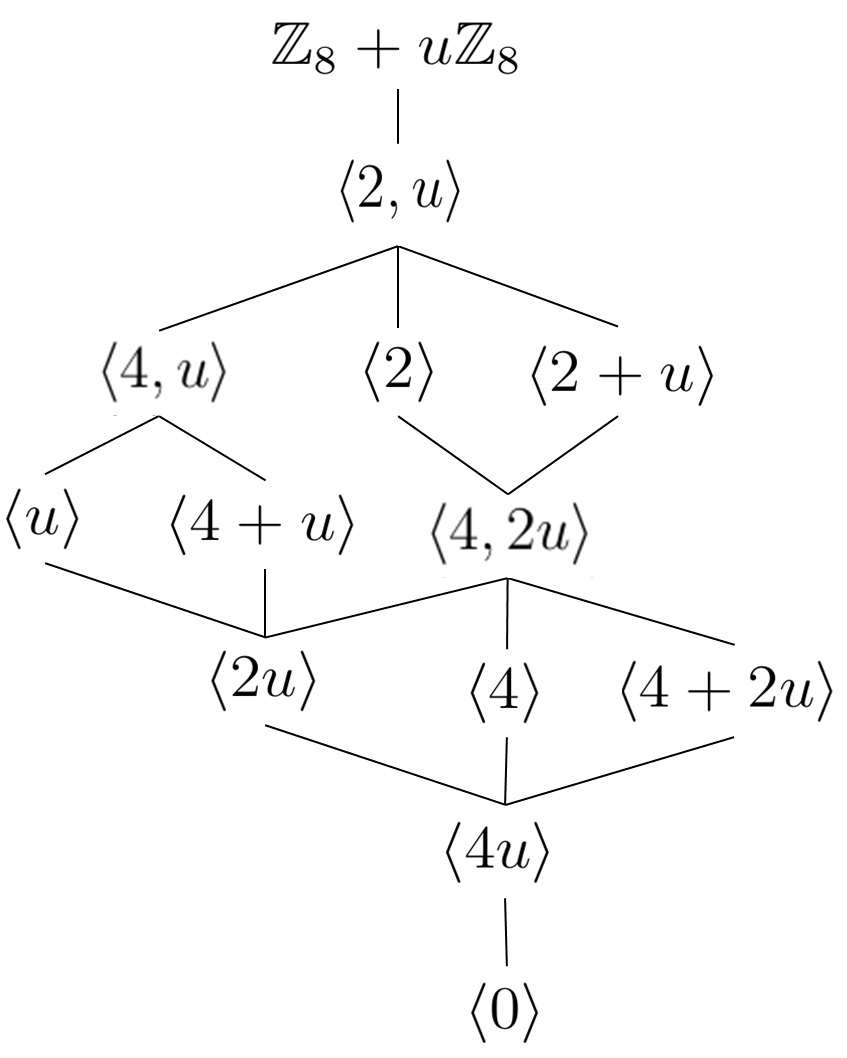}
%\resizebox{.25\textwidth}{!}{\includegraphics{z8.jpg} }
\caption{Lattice of Ideals of $\ZZ_8+u\ZZ_8$}
\label{fig:lat02}
\end{figure}

The ring $\ZZ_8+u\ZZ_8=\{ a+ub \mid a,b \in \ZZ_8, u^2=0 \}$ is a finite $64$-element commutative ring with identity of characteristic 8 which is a local, non-chain, and non-principal ideal ring. It behaves quite similarly with $\ZZ_4+u\ZZ_4$. The Jacobson radical is the unique maximal non-principal ideal $ \langle 2,u \rangle $ consisting of 32 elements of the form $a+ub$ where $a \in \{ 0,2,4,6\} $, the unique maximal ideal of the chain ring $\ZZ_8$. The group of units of $\ZZ_8+u\ZZ_8$ consists of 32 elements of the form $a+ub$ where $a \in \{ 1,3,5,7\} $, the group of units of $\ZZ_8$. The lattice in Fig. \ref{fig:lat02} shows the unique minimal principal ideal $\langle 4u \rangle = \{0,4u\}$ which is the socle. Consequently this ring is Frobenius and its generating character is given by $\chi(a+ub) = e^{\frac{\pi i}{4}(a+b)}$. It can also be shown that $\ZZ_8+u\ZZ_8$ is isomorphic to the quotient ring $\ZZ_8[x]/\langle x^2 \rangle$.   

We extend the Lee weight $w_L$ on $\ZZ_8$ given by $w_L(x) = \min{\{x,8-x}\}$ to $\ZZ_8+u\ZZ_8$ by $$w_L(a+ub) =w_L(b)+w_L(a+b).$$ As usual, this weight is not homogeneous. The homogeneous weight is derived analogously.

\begin{corollary}
The homogeneous weight on $\ZZ_8+u\ZZ_8$ is given by
\begin{equation}\label{hom8} w_{hom}(x) = \begin{cases}
                         0 & {\rm if} \ x=0 \\
                         2\Gamma & {\rm if }\ x=4u\\
												\Gamma & {\rm otherwise}\\
                    \end{cases}
\end{equation}
\end{corollary} 
Again observe that the highest possible homogeneous weight is given by the nonzero element of the minimal ideal, while all the other nonzero elements take the average value $\Gamma$ as their common weight. We obtain the following similar results.
\begin{theorem}
There is a weight preserving linear map from $(\ZZ_8+u\ZZ_8, \text{Lee weight})$ to $(\ZZ_8^2, \text{Lee weight})$.
\end{theorem}
\begin{theorem}
There is a weight preserving non-linear map $\phi_5$ from  $(\ZZ_8+u\ZZ_8, \text{Homogeneous weight})$ to $(\FF_2^{32}, \text{Hamming weight})$.
\end{theorem}

The map $\phi_5$ utilizes the generalized Gray isometry $\gamma_3$ from $\ZZ_{64}$ to $\FF_2^{16}$ whose image is the Boolean function on $\FF_2^5$ defined by $(y_1,y_2,y_3,y_4,y_5) \mapsto x_4+x_1y_1+x_2y_2+x_3y_3+x_4y_4+x_5y_5$.
   
\begin{theorem}
Let $B$ be a linear block code over  $\ZZ_8+u\ZZ_8$ of length $n$, then the associated image $\phi_5(B)$ is a (possibly nonlinear) binary block code of length $32n$. 
\end{theorem} 

\section{Summary and Conclusion}
\label{sect:con}
The method is evidently possible for the class of finite commutative rings $\ZZ_{2^r} + u\ZZ_{2^r}= \{ a+ub \mid a,b \in \ZZ_{2^r}, u^2=0 \}$ with $4^r$ elements and characteristic $2^r$. This ring is local, non-principal and a non-chain ring. The Jacobson radical is the unique maximal ideal $\langle 2,u \rangle$ consisting of $2^{2r-1}$ elements of the form $a+ub$ where $a$ is in the unique maximal ideal $\langle 2 \rangle$ with $2^{r-1}$ elements of the chain ring $\ZZ_{2^r}$. Thus $\abs{\ZZ_{2^r} + u\ZZ_{2^r}/\langle 2,u \rangle}=2$. The group of units of $\ZZ_{2^r}+u\ZZ_{2^r}$ is the difference $\ZZ_{2^r} + u\ZZ_{2^r} \setminus\langle 2,u \rangle$ which has cardinality $2^{2r-1}$ and takes the form $a+ub$ where $a$ is a unit in $\ZZ_{2^r}$. The socle is the unique minimal ideal $\langle 2^{r-1}u \rangle$ which clearly has only two elements $\{0, 2^{r-1} u\}$. Hence $\ZZ_{2^r} + u\ZZ_{2^r}$ is Frobenius and it induces a homogeneous weight similar to (\ref{hom4}) and (\ref{hom8}). To ensure that weights or distances are preserved, we compose certain isometries that capitalize on the generalized Gray map from $\ZZ_{2^r}$ to $\FF_2^{2^{r-1}}$ of Carlet \cite{car} which he defined as follows: let $x \in \ZZ_{2^r}$ and $x= \sum_{i=1}^{r} 2^{i-1}x_i$ as its binary expansion, $x_i \in \FF_2$. The image of $x$ by the generalized Gray map is the Boolean function $G$ on $\FF_2^{r-1}$ defined by $G(x):(y_1,y_2, \cdots,y_{r-1}) \rightarrow x_r + \sum_{i=1}^{r-1} x_i y_i.$ It will also be interesting to consider the linearity conditions on the binary image code. 

\section*{Acknowledgment}   

The first author acknowledges travel support from the UP System Research Dissemination Grant and the UPLB Academic Development Fund.

\end{document}